\documentclass[aps,prc,preprint,showpacs,showkeys,superscriptaddress,amsmath,amssymb,floatfix,nofootinbib]{revtex4}

\usepackage[dvips]{graphicx,color}
\usepackage{color}
\usepackage{dcolumn}
\usepackage{bm}
\usepackage{dcolumn}

\def\pmb#1{\setbox0=\hbox{#1}%
  \kern-.025em\copy0\kern-\wd0 
  \kern.05em\copy0\kern-\wd0
  \kern-.025em\raise.0433em\box0 }

\def\lambdabar{\protect\@lambdabar}
\def\@lambdabar{%
\relax
\bgroup
\def\@tempa{\hbox{\raise.73\ht0
\hbox to0pt{\kern.25\wd0\vrule width.5\wd0
height.1pt depth.1pt\hss}\box0}}%
\mathchoice{\setbox0\hbox{$\displaystyle\lambda$}\@tempa}%
{\setbox0\hbox{$\textstyle\lambda$}\@tempa}%
{\setbox0\hbox{$\scriptstyle\lambda$}\@tempa}%
{\setbox0\hbox{$\scriptscriptstyle\lambda$}\@tempa}%
\egroup
}

\begin{document}

\preprint{J-PARC-TH-0242}

\title{\boldmath
Production cross sections of $^{3,4}_\Lambda$H bound states in $^{3,4}$He($K^-$,~$\pi^0$) reactions at 1 GeV/$c$ 
}

\author{Toru~Harada}%
\email{harada@osakac.ac.jp}
\affiliation{%
Center for Physics and Mathematics,
Osaka Electro-Communication University, Neyagawa, Osaka, 572-8530, Japan
}
\affiliation{%
J-PARC Branch, KEK Theory Center, Institute of Particle and Nuclear Studies,
High Energy Accelerator Research Organization (KEK),
203-1, Shirakata, Tokai, Ibaraki, 319-1106, Japan
}

\author{Yoshiharu~Hirabayashi}%
\affiliation{%
Information Initiative Center, 
Hokkaido University, Sapporo, 060-0811, Japan
}

\date{\today}

\begin{abstract}
We investigate theoretically productions of $^{3,4}_\Lambda$H bound states
in the exothermic ($K^-$,~$\pi^0$) reactions on $^{3,4}$He targets at $p_{K^-}=$ 1.0 GeV/c 
in a distorted-wave impulse approximation with the optimal Fermi-averaging $K^-p\to\pi^0\Lambda$ $t$ matrix. 
We calculate angular distributions of the laboratory differential cross sections
 $d\sigma/d\Omega_{\rm lab}$ 
and the integrated cross sections $\sigma_{\rm lab}$ for a $J^P= 0^+$ ground state and 
a $J^P=1^+$ excited state of ${^{4}_\Lambda{\rm H}}$
at $\pi^0$ forward-direction angles of $\theta_{\rm lab}=$ 0$^\circ$--20$^\circ$, 
and those for a $J^P= 1/2^+$ ground state of ${^{3}_\Lambda{\rm H}}$.
The production of a $J^P=3/2^+$ excited state of ${^{3}_\Lambda{\rm H}}$ 
as a virtual state is also evaluated. 
The comparison in $d\sigma/d\Omega_{\rm lab}$ and $\sigma_{\rm lab}$ 
between $^{4}_\Lambda{\rm H}$ and $^{3}_\Lambda{\rm H}$
provides examining the mechanism of the production and structure of $^{3,4}_\Lambda$H,
as well as in the endothermic ($\pi^-$,~$K^0$) reactions at $p_{\pi^-}=$ 1.05 GeV/c.
This investigation confirms the feasibility of lifetime measurements 
of $^3_\Lambda$H at the J-PARC experiments. 
\end{abstract}
\maketitle


\section{Introduction}

Recently, experimental measurements of a ${^{3}_\Lambda{\rm H}}$ lifetime are planned  
by ($K^-$,~$\pi^0$) and ($\pi^-$,~$K^0$) reactions on a $^{3}$He target 
at J-PARC~\cite{J-PARC-P73,J-PARC-P74} 
to solve the puzzle that the unexpected short lifetime of ${^{3}_\Lambda{\rm H}}$ 
was measured in hypernuclear production by high-energy heavy-ion collisions \cite{HypHI13,Gal19}.
It seems that it is rather difficult to form a $\Lambda$ bound state by the 
nuclear ($K^-$,~$\pi^0$) and ($\pi^-$,~$K^0$) reactions 
because the $\Lambda$ hyperon is very weakly bound in $^3_\Lambda$H with a 
$J^P=$ ${1/2}^+$ ground-state (g.s.) separation energy $B_\Lambda=$ $0.13 \pm 0.05$ MeV \cite{Juric73}
with respect to the $d$-$\Lambda$ threshold, whereas a recent measurement by the STAR Collaboration 
reports a value of  $0.41 \pm 0.12$ MeV \cite{Adam20}. 

On the other hand, the production of a $J^P=$ ${0}^+$ ground state of $^4_\Lambda$He 
in the $^4$He($K^-$,~$\pi^-$) reaction 
is accomplished theoretically \cite{Harada98,Harada15} and experimentally \cite{Nagae98,Yamamoto15},
where the $\Lambda$ is bound with $B_\Lambda=$ $2.39 \pm 0.05$ MeV with respect 
to the $^3$He-$\Lambda$ threshold. 
In a previous paper \cite{Harada19}, we reexamined the production cross section of the 
$0^+_{\rm g.s.}$ state of $^4_\Lambda$He in the $^4$He($\pi$,~$K$) reaction at $p_{\pi^-}=$ 1.05 GeV/$c$, 
and discussed a benefit of the use of a $s$-shell target nucleus for $\Lambda$ production 
of the $A=$ 4 hypernucleus. 
To study the feasibility of the lifetime measurements of $^{3}_\Lambda$H in the production followed 
by mesonic decay processes~\cite{J-PARC-P73}, 
thus, it is worth examining theoretically the $\Lambda$ production of the $A=$ 3 hypernucleus 
via the ($\bar{K}$,~$\pi$) or ($\pi$,~$K$) reaction.

It has been recently discussed \cite{Schafer20} 
that there is a $s$-wave virtual state with $J^p=$ $3/2^+$, $L=0$ near the $d$-$\Lambda$ threshold 
in the $d + \Lambda$ system, which may correspond to a $J^P= 3/2^+$ excited state (exc) of $^{3}_\Lambda$H that 
has not yet been observed experimentally. 
Thus the production of the $3/2^+_{\rm exc}$ state of ${^{3}_\Lambda{\rm H}}$ via the 
$^3$He($K^-$,~$\pi^0$) reaction also needs to be examined from a theoretical point of view.

In this paper, we investigate theoretically the productions of the $^{3,4}_\Lambda$H bound states
in the exothermic ($K^-$,~$\pi^0$) reactions on $^{3,4}$He targets at $p_{K^-}=$ 1.0 GeV/$c$ 
in a distorted-wave impulse approximation (DWIA) using the optimal Fermi-averaging 
$K^-p\to\Lambda\pi^0$ $t$ matrix \cite{Harada05}.
We demonstrate angular distributions of the laboratory differential cross sections for 
${^{4}_\Lambda{\rm H}}$\,($J^P=0^+$, g.s.), ${^{4}_\Lambda{\rm H}}$\,($J^P=1^+$, exc), 
and ${^{3}_\Lambda{\rm H}}$\,($J^P=1/2^+$, g.s.) bound states 
in the $\pi^0$ forward-direction angles of $\theta_{\rm lab}=$ 0$^\circ$--20$^\circ$, 
and the integrated cross sections for them.
We also investigate the production of ${^{3}_\Lambda{\rm H}}$\,($J^P=3/2^+$, exc) as a virtual state 
close to the $d$-$\Lambda$ threshold. 
We discuss the effects of a weakly $\Lambda$ binding, a meson distortion, 
and an in-medium elementary amplitude 
in the nuclear ($K^-$,~$\pi^0$) reactions, 
as well as in the endothermic ($\pi^-$,~$K^0$) reaction~\cite{Harada19}.
To reduce uncertainties of several approximations and input parameters 
in our calculations, we attempt to examine the difference in the $\Lambda$ production 
between $^4_\Lambda$H and $^{3}_\Lambda$H.

\section{Calculations}

\subsection{Distorted-wave impulse approximation}

Let us consider a calculation procedure of the $\Lambda$ hypernuclear production 
for the nuclear ($K^-$,~$\pi$) reaction in the laboratory frame.
We will present briefly the standard DWIA calculation \cite{Hufner74,Dover80,Auerbach83}
applying to the productions of 
the $^{3,4}_\Lambda$H bound states in the reactions 
\begin{eqnarray}
K^- + {^{3,4}{\rm He}} \to \pi^0 + {^{3,4}_\Lambda{\rm H}}.  
\label{eqn:e1}
\end{eqnarray}
The differential cross section for the $\Lambda$ bound state with a spin-parity $J^P_B$ 
at the $\pi^0$ forward-direction angle of $\theta_{\rm lab}$ 
is written \cite{Harada19} by (in units $\hbar = c =1$)
\begin{eqnarray}
&&\left({d\sigma \over d\Omega}\right)_{{\rm lab}, \theta_{\rm lab}}^{J_B^P}
=  \alpha {1 \over {[J_A]}} \sum_{m_Am_B}
\biggl| \Bigl\langle {\Psi}_B 
\Big\vert\, \overline{f}_{\pi^0\Lambda}
          +i\,\overline{g}_{\pi^0\Lambda}\,{\bm \sigma}\cdot\hat{\bm n} \nonumber\\
&& \qquad \times \,
\chi^{(-)*}_{\pi}\left({\bm p}_{\pi},{M_{C} \over M_{B}}{\bm r}\right) 
\chi^{(+)}_{K^-}\left({\bm p}_{K^-},{M_{C} \over M_{A}}{\bm r}\right) 
\Big| \Psi_{A} \Bigr\rangle \biggr|^2, \nonumber\\
\label{eqn:e2}
\end{eqnarray}
where $[J]=2J+1$, and $\Psi_B$ and $\Psi_A$ are wave functions of the hypernuclear final state 
and the initial state of the target nucleus, respectively. 
The kinematical factor $\alpha$ denotes the translation from a two-body $K^-$-nucleon 
laboratory system to a $K^-$-nucleus laboratory system \cite{Dover83}.
$\hat{\bm n}$ is a unit vector perpendicular to the reaction plane. 
$\chi_{\pi}^{(-)}$ and $\chi_{K^-}^{(+)}$ are meson distorted waves for outgoing $\pi^0$ 
and incoming $K^-$, respectively. 
The factors of $M_{C}/M_{B}$ and $M_{C}/M_{A}$ arise from the recoil correction, where 
$M_{A}$, $M_{B}$ and $M_{C}$ are the masses of the target, the hypernucleus, and the core nucleus, respectively.
The energy and momentum transfers to the final state are given by
\begin{eqnarray}
&\omega = E_{K^-}-E_\pi, &\quad {\bm q} ={\bm p}_{K^-}-{\bm p}_{\pi}, 
\label{eqn:e3}
\end{eqnarray}
where $E_{K^-}=({\bm p}_{K^-}^2+m_{K^-}^2)^{1/2}$ and $E_\pi=({\bm p}_\pi^2+m_\pi^2)^{1/2}$ 
(${\bm p}_{K^-}$ and ${\bm p}_{\pi}$) are the laboratory energies (momenta) 
of $K^-$ and $\pi^0$ in this nuclear reaction, respectively; $m_{K^-}$ and $m_\pi$ are
the masses of $K^-$ and $\pi^0$, respectively.
$\overline{f}_{\pi^0\Lambda}$ and $\overline{g}_{\pi^0\Lambda}$ describe 
the non-spin-flip $\Delta S=$ 0 and spin-flip $\Delta S=$ 1 amplitudes for the 
in-medium $K^-p\to \pi^0 \Lambda$ production, respectively, 
which take into account the Fermi motion of a struck nucleon in the nuclear target 
for the nuclear ($K^-$,~$\pi^0$) reaction.
The explicit forms of Eq.~(\ref{eqn:e1}) are given by Appendix \ref{app:1}. 

The Fermi-averaging treatment may essentially affect the absolute values of the production 
cross sections even through $^{3}_\Lambda$H \cite{Mart08}. 
Here we apply the optimal Fermi-averaging procedure \cite{Harada05} to 
$\overline{f}_{\pi^0\Lambda}$ and $\overline{g}_{\pi^0\Lambda}$ in the nuclear ($K^-$,~$\pi^0$) reaction;
the momentum distribution $\rho(p)$ of the struck nucleon in $^3$He ($^4$He) is assumed as a simple harmonic
oscillator with a size parameter $b_N$ = 1.61 (1.33) fm, leading to 
$\langle p^2 \rangle^{1/2}\simeq$ 150 (182) MeV/$c$ in the nucleus~\cite{Harada14}. 
We employ the elementary $\bar{K}N\to \pi\Lambda$ amplitude analyzed by Gopal, et al.~\cite{Gopal77}.

The distorted waves of $\chi_{\pi}^{(-)}$ and $\chi_{K^-}^{(+)}$
are obtained in a computational procedure simplified with the help of 
the eikonal approximation \cite{Hufner74,Dover80}.
We choose $\alpha_{K^-}= \alpha_\pi =$ 0, $\sigma_{K^-}$= 45 mb, 
and $\sigma_\pi$= 32 mb in charge independence, 
as eikonal distortion parameters for the $^{3,4}$He targets.
Although such distortions are seemed to be not so important in 
the light $s$-shell nuclear systems than in $p$-shell nuclear systems like $^{12}$C, 
it is necessary to verify their effects on the production cross sections 
in more quantitative calculations \cite{Harada19}.

\begin{table*}[bt]
\caption{\label{tab:1}
Calculated $\Lambda$ separation energies $B_{\Lambda}$ 
and root-mean-square distances $\langle r^2_{\Lambda} \rangle^{1/2}$
between the core nucleus and $\Lambda$ in 
$^{3,4}_\Lambda{\rm H}$  with an isospin $T$ and a spin $J^P$, 
in comparison with $B_{N}$ and $\langle r^2_{N} \rangle^{1/2}$ for 
a nucleon in $^{3,4}{\rm He}$ targets. 
}
\begin{ruledtabular}
\begin{tabular}{llllllll}
&   \multicolumn{3}{c}{$A=3$} 
&&   \multicolumn{3}{c}{$A=4$} \\
\noalign{\smallskip}
     \cline{2-4} \cline{6-8}
\noalign{\smallskip}
&  \multicolumn{1}{l}{${^3_\Lambda{\rm H}_{\rm g.s.}}$}
&  \multicolumn{1}{l}{${^3_\Lambda{\rm H}_{\rm exc}}$}
&  \multicolumn{1}{l}{${^3{\rm He}}$}
&
&  \multicolumn{1}{l}{${^4_\Lambda{\rm H}_{\rm g.s.}}$}
&  \multicolumn{1}{l}{${^4_\Lambda{\rm H}_{\rm exc}}$}
&  \multicolumn{1}{l}{${^4{\rm He}}$} \\
\noalign{\smallskip}\hline\noalign{\smallskip}
$T$            &  $0$      & $0$      &  $1/2$  
               &&  $1/2$   & $1/2$    &  $0$ \\
$J^P$          &  $1/2^+$  & $3/2^+$  &  $1/2^+$  
               &&  $0^+$   & $1^+$    &  $0^+$ \\
$B_{\Lambda(N)}$ (MeV)  
               &  0.13  &  unbound\tablenotemark[1]    &  5.49  &&  2.16    &  1.09    &   19.8  \\
$\langle r^2_{\Lambda(N)} \rangle^{1/2}$ (fm)  
               &  11.2 &  18.2\tablenotemark[2]   & 2.49   &&   3.68   &  4.60    &   1.87   \\
$B^{\rm exp}_{\Lambda}$ (MeV)  
               &  $0.13\pm0.05$ \cite{Juric73}   &                                &   
               &&  $2.04\pm0.04$ \cite{Juric73}  &  $0.95\pm0.04$ \cite{Bedjidian76}  &   \\
               &  $0.41\pm0.12$ \cite{Adam20}    &                                &   
               && $2.16\pm0.08$ \cite{Schulz16} &  $1.09\pm0.02$ \cite{Yamamoto15} &   \\
\end{tabular}
\end{ruledtabular}
\tablenotetext[1]{
The pole of the $S$ matrix as the virtual state is located at $E_{\Lambda d}^{\rm (pole)}=$ 
$-$0.089 MeV on the unphysical sheet [$-$].} 
\tablenotetext[2]{The continuum-discretized wave function is used. See text in Sect.~\ref{virtual}.}
\end{table*}

\begin{figure}[t]
\begin{center}
  \includegraphics[width=\linewidth]{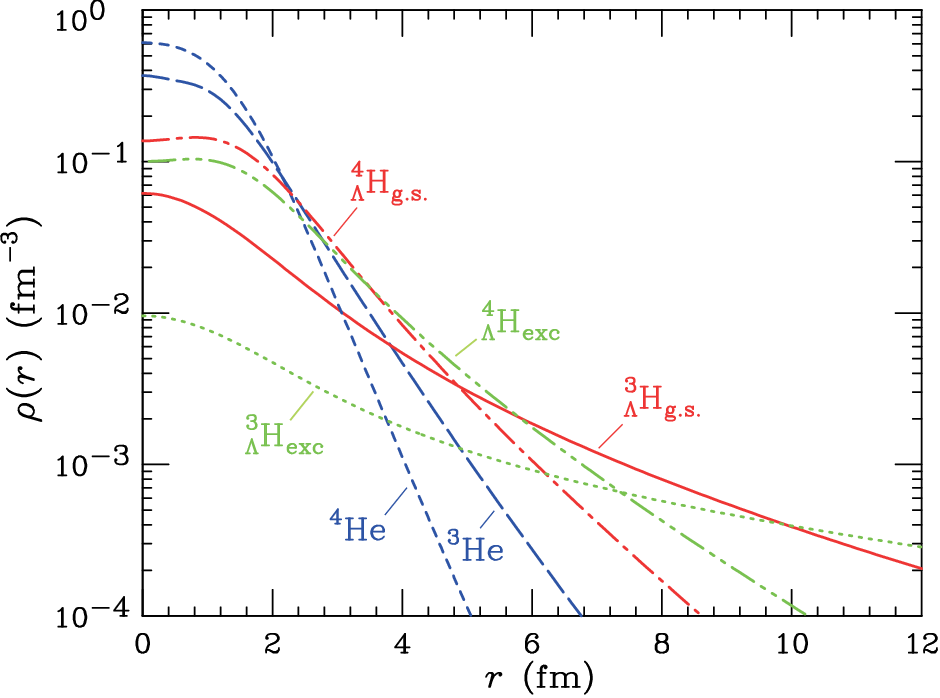}
\end{center}
\caption{\label{fig:1}
Relative density distributions $\rho_\Lambda(r)$ for $\Lambda$ in ${^{3,4}_\Lambda{\rm H}_{\rm g.s.}}$, 
as a function of the relative distance between the core nucleus and $\Lambda$, 
together with the relative density distributions $\rho_N(r)$ for a nucleon ($N$) 
in the $^{3,4}{\rm He}$ targets. 
The relative density distributions for $\Lambda$ in ${^{3,4}_\Lambda{\rm H}_{\rm exc}}$ are 
also drawn. 
}
\end{figure}

\subsection{\boldmath
Effective number of nucleons for $^{3,4}_{\Lambda}{\rm H}$ bound-state productions}

Considering the non-spin-flip $\Delta S=$ 0 production in the $^{3,4}$He($K^-$, $\pi^0$) reactions, 
we obtain the differential cross sections of Eq.~(\ref{eqn:e2}) 
for the production of ${^{3,4}_\Lambda{\rm H}}$, 
which are often written by the effective number technique \cite{Hufner74,Dover80};
\begin{equation}
\left({d\sigma \over d\Omega}\right)_{\rm lab, \theta_{\rm lab}}^{J_B^P}
= \alpha \left\langle{\frac{d\sigma}{d\Omega}}\right\rangle^{\rm elem}_{\rm lab, \theta_{\rm lab}} 
Z^{J_B^P}_{\rm eff}(\theta_{\rm lab}),
\label{eqn:e4}
\end{equation}
where $\alpha\langle{d\sigma/d\Omega}\rangle^{\rm elem}_{\rm lab}=\alpha |\overline{f}_{\pi^0\Lambda}|^2$
is a differential cross section for the in-medium ${K^-}p\to \pi^0\Lambda$ $\Delta S=$ 0 reaction, 
including the kinematical factor $\alpha$. 
The effective number of nucleons $Z^{J_B^P}_{\rm eff}$ for the production of 
the $^{3,4}_\Lambda{\rm H}$\,($J^P_B$) bound state is reduced \cite{Harada19} as
\begin{eqnarray}
Z^{J_B^P}_{\rm eff}(\theta_{\rm lab})& = &  C^2_{TS}|F(q)|^2,
\label{eqn:e5}
\end{eqnarray}
where $C_{TS}$ is the isospin-spin ($TS$) spectroscopic amplitude between 
the $\Lambda$ final state and the $N$ initial state. 
The form factor $F(q)$ is written by 
\begin{eqnarray}
F(q)&=&
\int_0^{\infty} dr r^2
\rho_{\rm tr}(r)\widetilde{j}_{0}\left(q; \frac{M_C}{M_D}r\right),
\label{eqn:e6} 
\end{eqnarray}
where $\widetilde{j}_{0}(q; r)$ is a distorted wave 
for the meson distortion; $M_C/M_B$ and $M_C/M_A$ in Eq.~(\ref{eqn:e2}) are replaced 
by $M_C/M_D$ in the eikonal approximation, 
where an averaged mass $M_D \equiv (M_B+M_A)/2$, which may give 
a good estimation for the very light nuclear systems.
Because the factor of $M_C/M_D$ originates from the recoil correction, 
the {\it effective} momentum transfer is often defined by 
$q_{\rm eff}\equiv (M_C/M_D)q \simeq [(A-1)/A]q$, which controls effectively the recoil effects 
on the production cross sections in the eikonal approximation~\cite{Dover80,Auerbach83}. 
When the distortion effects are switched off ($\sigma_{K^-}, \sigma_\pi\to 0$), 
the production cross sections can be obtained in the plane-wave impulse approximation (PWIA),
with replacing $\tilde{j}_{0}(q; r)$ by ${j}_{0}(qr)$ that is a spherical 
Bessel function with $L=0$~\cite{Harada19}.

The transition density $\rho_{\rm tr}(r)$ in Eq.~(\ref{eqn:e6}) is given by   
\begin{eqnarray}
\rho_{\rm tr}(r)
&=&{\varphi}^{(\Lambda)*}_{0}(r)\varphi^{(N)}_{0}(r),
\label{eqn:e7}
\end{eqnarray}
where $\varphi^{(\Lambda)}_{0}=\langle\,\phi^{(C)}_0| \Psi_{B}\rangle$ is the relative 
wave function that is regarded as a spectroscopic amplitude for $\Lambda$ 
in ${^{3,4}_\Lambda{\rm H}}$ by using the wave function $\phi^{(C)}_0$ for 
the core nucleus~\cite{Akaishi86}, and $\varphi^{(N)}_{0}=\langle\,\phi^{(C)}_0| \Psi_{A}\rangle$ 
is the relative wave function for a nucleon ($N$) in ${^{3,4}{\rm He}}$. 
In Table~\ref{tab:1}, we show the calculated $\Lambda$ separation energies $B_\Lambda$ 
and the root-mean-square distances $\langle r_\Lambda^2 \rangle^{1/2}$ between the core nucleus 
and $\Lambda$ in ${^{3,4}_\Lambda{\rm H}}$ with the isospin $T_B$ and the spin $J^P_B$, 
together with the nucleon separation energies $B_N$ and $\langle r_N^2 \rangle^{1/2}$ 
between the core nucleus and $N$ in ${^{3,4}{\rm He}}$ with $T_A$ and $J^P_A$.
For $A=4$, we obtain $\varphi^{(\Lambda)}_{0}$ in the $3N$-$\Lambda$ model 
based on four-body $\Lambda NNN$ calculations~\cite{Harada19} with central nucleon-nucleon 
($NN$) and $\Lambda N$ potentials, 
reproducing $B_\Lambda =$ 2.16 MeV \cite{Schulz16} and $\langle r^2_\Lambda \rangle^{1/2}=$ 3.68 fm 
for $^3{\rm H} + \Lambda$ in ${^{4}_\Lambda{\rm H}}$\,($J^P=0^+$, g.s.). 
We use $\varphi^{(N)}_{0}$ obtained in four-body $NNNN$ calculations~\cite{Akaishi86}, 
and $C^2_{TS}=2$. 
For $A=3$, we obtain $\varphi^{(\Lambda)}_{0}$ in the $2N$-$\Lambda$ model based on 
microscopic continuum-discretized coupled-channels (CDCC) calculations \cite{Kaminura89,Harada15a} 
with central $NN$ and $\Lambda N$ potentials, 
reproducing $B_\Lambda =$ 0.13 MeV \cite{Juric73} and $\langle r^2_\Lambda \rangle^{1/2}=$ 11.2 fm 
for $d + \Lambda$ in ${^{3}_\Lambda{\rm H}}$\,($J^P=1/2^+$, g.s.).
We use $\varphi^{(N)}_{0}$ obtained in three-body $NNN$ calculations~\cite{Koike96}, and 
$C^2_{TS}=3/2$ in Eq.~(\ref{eqn:e5}).
Figure~\ref{fig:1} shows the relative density distributions 
$\rho_\Lambda(r)=|\varphi^{(\Lambda)}_{0}(r)|^2$ for $\Lambda$ in $^{3,4}_\Lambda{\rm H}_{\rm g.s.}$, 
as a function of the relative distance between the core nucleus and $\Lambda$, 
together with relative density distributions 
$\rho_N(r)=|\varphi^{(N)}_{0}(r)|^2$ for $N$ in ${^{3,4}{\rm He}}$.
We confirm that the $\Lambda$ density distribution for $^4_\Lambda{\rm H}_{\rm g.s.}$ is 
significantly suppressed at the nuclear center and is pushed outside~\cite{Kurihara85,Harada19}, 
and that the $\Lambda$ density distribution for $^3_\Lambda{\rm H}_{\rm g.s.}$ indicates 
the weakly bound state having a long tail.

Considering the spin-flip $\Delta S=$ 1 production in the $^{4}{\rm He}$($K^-$, $\pi^0$) reaction, 
we obtain the production cross section of an unnatural-parity state of 
${^4_\Lambda{\rm H}}$\,($J^P=1^+$, exc).
Due to such a $s$-shell hypernuclear state, the differential cross section 
is also written by the effective number of nucleons $Z^{J_B^P}_{\rm eff}$ with 
$\alpha\langle{d\sigma/d\Omega}\rangle^{\rm elem}_{\rm lab}= \alpha |\overline{g}_{\pi^0\Lambda}|^2$
for the in-medium ${K^-}p\to \pi^0\Lambda$ $\Delta S=$ 1 reaction, as given in Eq.~(\ref{eqn:e4}). 
We use $\varphi^{(\Lambda)}_{0}$ obtained by the $3N$-$\Lambda$ model 
with central $\Lambda N$ potentials for $J^P=1^+$, 
reproducing $B_\Lambda =$ 1.09 MeV with respect to the $^3{\rm H}$-$\Lambda$ threshold \cite{Yamamoto15} 
with $\langle r^2_\Lambda \rangle^{1/2}=$ 4.60 fm, and $C^2_{TS}=1$ in Eq.~(\ref{eqn:e5}).

\subsection{\boldmath\label{virtual}
${^{3}_\Lambda{\rm H}}$\,($J^P=3/2^+$, exc) as a virtual state
}

We have no observation of a $J^P= 3/2^+$ excited state of $^{3}_\Lambda{\rm H}$ experimentally so far.
It seems that there is no bound state in $J^P= 3/2^+$, but this state would be 
in the continuum region above the $d$-$\Lambda$ threshold \cite{Mart08} as a resonant state 
or a virtual state theoretically.
If a pole of the $S$ matrix for a virtual state is sufficiently close to the physical axis 
on the complex momentum plane, it may provide an appreciable influence on the cross section 
at low energy \cite{Taylor06}.
This phenomenon is regarded as {\it threshold effects} caused by a virtual state 
close to the threshold.
To see this situation, we study the $s$-wave $3/2^+$ state in $^{3}_\Lambda{\rm H}$ 
with a folding model $d$-$\Lambda$ potential $U_{\Lambda d}$, adjusting 
to the $\Lambda d$ scattering length of $a_{3/2}^{\Lambda d}=$ $-$16.2 fm 
and the effective range of $r_{3/2}^{\Lambda d}=$ 3.2 fm \cite{Schafer20} for the NSC97f potential.
Numerically solving the Lippmann-Schwinger equation for this $d+\Lambda$ system, 
we find that its pole of the $S$ matrix is located at $k^{\rm (pole)}_{\Lambda d}=$ $-0.057i$ fm$^{-1}$ 
on the complex momentum plane, which corresponds to $E^{\rm (pole)}_{\Lambda d}=$ $-$0.089 MeV 
on the unphysical sheet [$-$] of the complex $E$ plane \cite{Harada17}. 
This state is identified as a virtual state close to the $d$-$\Lambda$ threshold, as recently discussed by 
Sch\"{a}fer et al.~\cite{Schafer20}.
To describe the virtual state of ${^{3}_\Lambda{\rm H}}$\,($J^P=3/2^+$, exc), 
we construct a continuum-discretized wave function $\hat{\varphi}^{(\Lambda)}_{0}$, 
which is given by an appropriate momentum bin of $k_0$ and $k_1$ \cite{Kawai86}:
\begin{eqnarray}
&& \hat{\varphi}^{(\Lambda)}_{0}({\bm r})
={1 \over \sqrt{\Delta k}}\int^{k_1}_{k_0}\phi_0^{d+\Lambda}(k,{\bm r})dk, 
\label{eqn:e8}
\end{eqnarray}
where $\Delta k=k_1-k_0$, and ${\bm r}$ and $k$ are the radial coordinate and the relative momentum 
between $d$ and $\Lambda$, respectively, and 
$\phi_0^{d+\Lambda}(k,{\bm r})$ is a $d + \Lambda$ scattering wave function with the energy 
$\varepsilon_{\Lambda d}=k^2/(2\mu_{\Lambda d})$ $(> 0)$, where $\mu_{\Lambda d}$ is the reduced mass 
of the $d + \Lambda$ system.
This continuum-discretized wave function $\hat{\varphi}^{(\Lambda)}_{0}$ satisfies the {\it positive} energy 
$\hat{\varepsilon}_{\Lambda d}=\{(k_1+k_0)^2/4+(\Delta k)^2/12\}/(2\mu_{\Lambda d})$ \cite{Kawai86}.
When we choose the momentum bin of $k_0=$ 0.026 fm$^{-1}$ and $k_1 =$ 0.083 fm$^{-1}$ for ${^{3}_\Lambda{\rm H}}$\,($J^P=3/2^+$, exc), 
we find that the wave functions of $\phi_0^{d+\Lambda}(k,{\bm r})$ in Eq.~(\ref{eqn:e8}) 
significantly enhance at the nuclear interior
due to the threshold effects caused by the virtual state \cite{McVoy68}. 
Thus, the continuum-discretized wave function $\hat{\varphi}^{(\Lambda)}_{0}$ labeled by
$\hat{\varepsilon}_{\Lambda d}=$ 0.089 MeV provides an appropriate description of these threshold effects, 
consistent with the virtual-state phenomenon having a peak around 
$\hat{\varepsilon}_{\Lambda d}\simeq |E_{\Lambda d}^{\rm (pole)}|$ 
above the threshold in the nuclear response function \cite{Morimatsu94}. 
Regarding $\hat{\varphi}^{(\Lambda)}_{0}$ as $\varphi^{(\Lambda)}_{0}$ in Eq.~(\ref{eqn:e7}), 
therefore, we can estimate the production cross sections of ${^{3}_\Lambda{\rm H}}$\,($J^P=3/2^+$, exc)
as the virtual state with $C^2_{TS}=$ 4/3, as given in Eq.~(\ref{eqn:e5}). 
In Fig.~\ref{fig:1}, we also display the relative density distribution $\rho_\Lambda(r)$ 
for ${^{3}_\Lambda{\rm H}}$\,($J^P=3/2^+$, exc), which leads to 
$\langle r^2_\Lambda \rangle^{1/2}=$ 18.2 fm, 
in comparison with ${^{3}_\Lambda{\rm H}}$\,($J^P=1/2^+$, g.s.).

\section{Results and discussion}

\subsection{\boldmath
Optimal Fermi-averaged differential cross sections
}

It has been recognized that in the standard DWIA, 
the in-medium $K^-p\to \pi^0\Lambda$ cross section of 
$\alpha\langle{d\sigma/d\Omega}\rangle^{\rm elem}_{\rm lab}$
in Eq.~(\ref{eqn:e4}) plays an important role in explaining the production cross 
section of $d\sigma/d\Omega_{\rm lab}$ in the nuclear reaction.
To realize a more quantitative description in our calculations, 
we need to obtain the in-medium $K^-p\to \pi^0\Lambda$ cross section with 
the optimal Fermi-averaging $t$ matrix \cite{Harada05} 
for the nuclear ($K^-$,~$\pi^0$) reaction, taking into account momenta arising from 
the distorted waves of $K^-$ and $\pi^0$ in the nucleus.

\begin{figure}[t]
\begin{center}
  \includegraphics[width=\linewidth]{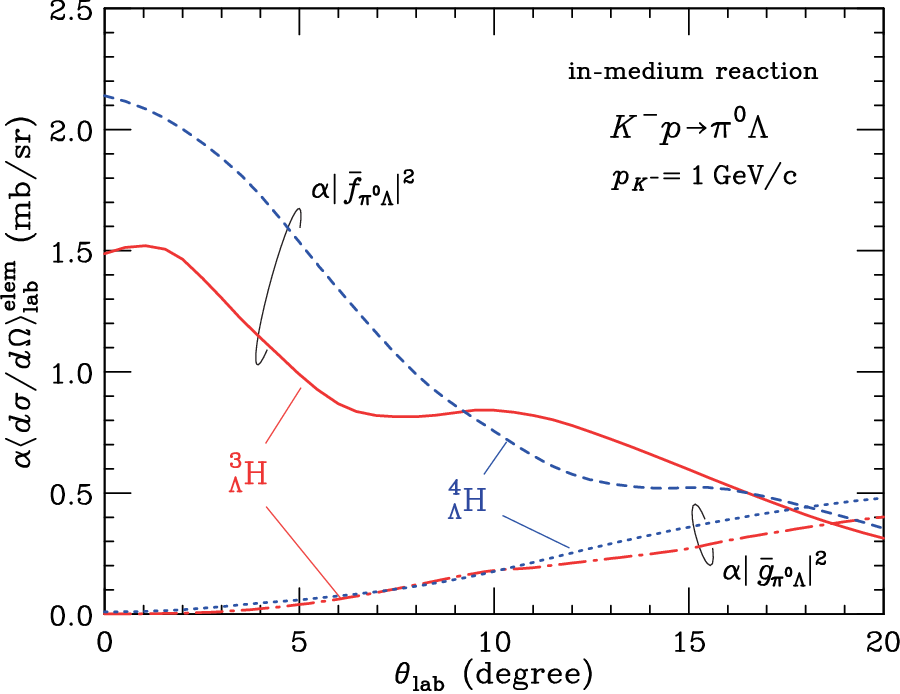}
\end{center}
\caption{\label{fig:2}
Angular distributions of the in-medium $K^-p\to \pi^0\Lambda$ 
differential cross sections 
$\alpha\langle{d\sigma/d\Omega}\rangle^{\rm elem}_{\rm lab}=\alpha|\overline{f}_{\pi^0\Lambda}|^2$ 
for the non-spin-flip $\Delta S=$ 0 production in the ($K^-$, $\pi^0$) reactions at 
$p_{K^-}=$ 1.0 GeV/$c$, and those of $\alpha |\overline{g}_{\pi^0\Lambda}|^2$ for 
the spin-flip $\Delta S=$ 1 production. 
Solid and dot-dashed (dashed and dotted)  
curves denote the calculated results for ${^3_\Lambda{\rm H}}$ 
(${^4_\Lambda{\rm H}}$) production on the $^3$He ($^4$He) target, respectively.
The optimal Fermi-averaging $K^-p\to \pi^0\Lambda$ amplitudes of 
$\overline{f}_{\pi^0\Lambda}$ and $\overline{g}_{\pi^0\Lambda}$ 
are obtained \cite{Harada05} in the use of the elementary amplitudes analyzed 
by Gopal, et al.~\cite{Gopal77}.
}
\end{figure}

Figure~\ref{fig:2} displays the angular distributions of the in-medium $K^-p\to \pi^0\Lambda$
differential cross sections for non-spin-flip $\Delta S=$ 0 and spin-flip $\Delta S=$ 1 productions, 
which are denoted by $\alpha\langle{d\sigma/d\Omega}\rangle^{\rm elem}_{\rm lab}= 
\alpha |\overline{f}_{\pi^0\Lambda}|^2$ and $\alpha |\overline{g}_{\pi^0\Lambda}|^2$, respectively, 
in the ($K^-$, $\pi^0$) reactions on $^{3,4}$He at $p_{K^-}=$ 1.0 GeV/$c$.
These behaviors affect the angular distributions of 
the productions of ${^{3,4}_\Lambda{\rm H}}$, as well as the behavior of $Z_{\rm eff}$ that 
depends on the angle of $\theta_{\rm lab}$.
Note that the absolute values and shapes of 
$\alpha|\overline{f}_{\pi^0\Lambda}|^2$ on $^{4}$He and $^{3}$He 
differ moderately in the $\Delta S=0$ productions. 
This difference stems from the nature of the energy and momentum transfers ($\omega$, ${\bm q}$) 
that satisfy the on-shell energy condition in the optimal Fermi-averaging procedure
\begin{eqnarray}
\omega
&\simeq& m_\Lambda-m_N+{{\bm q}^2 \over {2m_\Lambda}}
   +{{{\bm p}^*_N}\cdot{\bm q} \over {m_\Lambda}}
   -{{m_\Lambda-m_N} \over m_\Lambda} { {{\bm p}^*_N}^{2} \over 2 m_N} \nonumber \\
&=& m_\Lambda - m_N -B_\Lambda +B_N + T_{\rm recoil},
\label{eqn:e7a} 
\end{eqnarray}
where ${\bm p}^*_N$ is the momentum of a struck nucleon 
and $T_{\rm recoil}$ is the recoil energy to the final state.
Because the nucleon binding energy of $B_N=$ 19.8 MeV for $^4$He is so larger than 
that of $B_N=$ 5.49 MeV for $^3$He, 
the energy difference $\Delta\omega =|\omega_4-\omega_3|$ 
becomes $\sim$11 MeV, where 
$\omega_{4}$ and $\omega_{3}$ are the energy transfers to $^{4}_\Lambda$H and $^{3}_\Lambda$H, respectively. 
This value is comparable to ${\bm q}^2/2m_\Lambda \simeq$ 3.6--12 MeV 
in the near-recoilless ($K^-$, $\pi^0$) reactions at $\theta_{\rm lab}=$ 0$^\circ$--8$^\circ$. 
As a result, it induces a significant downward energy shift to the $\Lambda$ production threshold 
in the nuclear ($K^-$,$\pi^0$) reactions on $^4$He, rather than $^3$He.
This fact leads to the difference in $\alpha |\overline{f}_{\pi^0\Lambda}|^2$ 
between $^4$He and $^3$He owing to the energy dependence of 
the elementary amplitude $f_{\pi^0\Lambda}$ at the forward angles of $\theta_{\rm lab}$.
On the other hand, we find that the values of $\alpha |\overline{g}_{\pi^0\Lambda}|^2$ 
in the $\Delta S=1$ productions on $^{4}$He and $^{3}$He are very similar.

\subsection{\boldmath
${^{4}_\Lambda{\rm H}}$\,($J=0^+$, g.s.) and ${^{4}_\Lambda{\rm H}}$\,($J^P=1^+$, exc)
}

\begin{figure}[t]
\begin{center}
  \includegraphics[width=\linewidth]{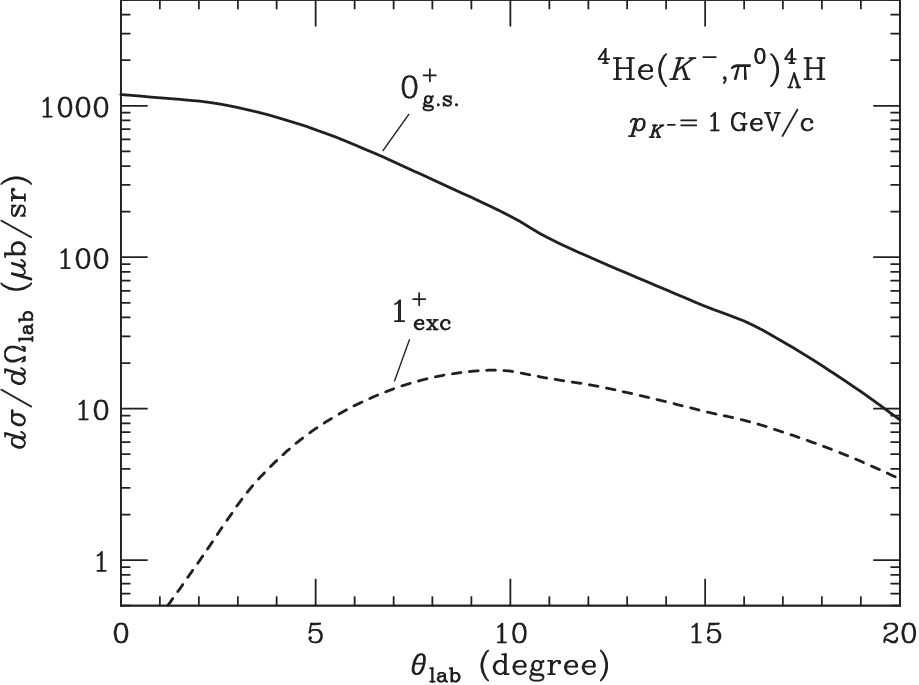}
\end{center}
\caption{\label{fig:3}
Calculated angular distributions of the laboratory differential cross sections 
$d\sigma/d\Omega_{\rm lab}$ for $0^+_{\rm g.s.}$ and $1^+_{\rm exc}$ states 
in $^{4}_\Lambda$H via the $^{4}$He($K^-$,~$\pi^0$) 
reactions at $p_{K^-}=$ 1.0 GeV/$c$ in the DWIA.
}
\end{figure}

Now we estimate numerically the production cross sections of $^{4}_\Lambda$H 
in the exothermic $^{4}$He($K^-$,~$\pi^0$) reactions at $p_{K^-}=$ 1.0 GeV/$c$ 
and $\theta_{\rm lab}=$ 0$^\circ$--20$^\circ$, where the momentum transfers 
become $q \simeq$ 90--345 MeV/$c$. 
In Table~\ref{tab:2}, we list the calculated results of 
$\alpha\langle{d\sigma/d\Omega}\rangle^{\rm elem}_{\rm lab}$, $Z_{\rm eff}$,
and $d\sigma/d\Omega_{\rm lab}$ for 
${^{4}_\Lambda{\rm H}}$\,($J=0^+$, g.s.) and ${^{4}_\Lambda{\rm H}}$\,($J^P=1^+$, exc)
in the DWIA using the distortion parameters 
of ($\sigma_{K^-}$,~$\sigma_\pi$) = (45 mb, 32 mb).
We obtain 
$d\sigma/d\Omega_{\rm lab}(0^+_{\rm g.s.})=$ 1184.3, 552.2, 100.8, and 19.2 $\mu$b/sr
and 
$d\sigma/d\Omega_{\rm lab}(1^+_{\rm exc})=$ 0.27, 10.5, 14.4, and 5.68 $\mu$b/sr
at $\theta_{\rm lab}=$ 0$^{\circ}$, 6$^{\circ}$, 12$^{\circ}$, and 18$^{\circ}$, respectively. 
In Fig.~\ref{fig:3}, 
we show the calculated angular distributions of 
$d\sigma/d\Omega_{\rm lab}$ for the $0^+_{\rm g.s.}$ and $1^+_{\rm exc}$ states 
in $^{4}_\Lambda$H via the $^{4}$He($K^-$,~$\pi^0$) reactions at $p_{K^-}=$ 1.0 GeV/$c$ in the DWIA.
We find that the production cross section of the $0^+_{\rm g.s.}$ state dominates in the forward angles, 
whereas the production of the $1^+_{\rm exc}$ state is comparable to that of the $0^+_{\rm g.s.}$ state 
beyond $\theta_{\rm lab}=$ 20$^\circ$; we have 
$[d\sigma/d\Omega_{\rm lab}(0^+_{\rm g.s.})]/
[d\sigma/d\Omega_{\rm lab}(1^+_{\rm exc})] \simeq$ 2.5
at $\theta_{\rm lab}\simeq$ 20$^{\circ}$.

The integrated cross section of $^{4}_\Lambda$H 
over $\theta_{\rm lab}=$ 0$^{\circ}$--20$^{\circ}$ is given by
\begin{eqnarray}
\sigma_{\rm lab}(J^P_B)\equiv
\int_{\theta_{\rm lab}= 0^\circ}^{\theta_{\rm lab}= 20^\circ} \!\!
\left(\frac{d\sigma}{d\Omega}\right)_{\rm lab, \theta_{\rm lab}}^{J_B^P} d\Omega.
\label{eqn:e9}
\end{eqnarray}
We find $\sigma_{\rm lab}(0^+_{\rm g.s.})=$ 63.1 $\mu$b 
and 
$\sigma_{\rm lab}(1^+_{\rm exc})=$ 3.8 $\mu$b 
at $p_{K^-}=$ 1.0 GeV/$c$ in the DWIA, 
compared with $\sigma_{\rm lab}(0^+_{\rm g.s.})=$ 150.4 $\mu$b 
and 
$\sigma_{\rm lab}(1^+_{\rm exc})=$ 10.7 $\mu$b in the PWIA.
This implies that distortion effects are remarkably important in 
quantitative estimations for the $A=$ 4 nuclear systems \cite{Harada19}. 
The distortion effects are roughly estimated by a distortion factor 
$D_{\rm dis} \equiv Z^{\rm DW}_{\rm eff}/Z^{\rm PW}_{\rm eff}$ in the eikonal meson waves.
We find 
$D_{\rm dis} \simeq$ 0.42--0.23 for $^4_\Lambda$H, which depend 
on $\theta_{\rm lab}=$ 0$^\circ$--20$^\circ$.
Consequently, we show that the production ratio of $\sigma_{\rm lab}(1^+_{\rm exc})$
to $\sigma_{\rm lab}(0^+_{\rm g.s.})$ amounts to 
\begin{eqnarray}
R_4=\sigma_{\rm lab}(1^+_{\rm exc})/\sigma_{\rm lab}(0^+_{\rm g.s.})
\simeq 0.06\mbox{--}0.07 \quad \mbox{for $^4_\Lambda{\rm H}$}
\label{eqn:e10}
\end{eqnarray}
at $p_{K^-}=$ 1.0 GeV/$c$.

\begin{table*}[bt]
\caption{\label{tab:2}
Calculated angular distributions of the laboratory differential cross sections for the 
$0^+_{\rm g.s.}$ and $1^+_{\rm exc}$ states in $^{4}_\Lambda$H via the $^{4}$He($K^-$,~$\pi^0$) 
reactions at $p_{K^-}=$ 1.0 GeV/$c$ in the DWIA. 
The distortion parameters of ($\sigma_{K^-}$,~$\sigma_\pi$) = (45 mb, 32 mb) are used.
}
\begin{ruledtabular}
\begin{tabular}{ccccrcccr}
&  
&   \multicolumn{3}{c}{${^4_\Lambda{\rm H}}$\,($J^P=0^+$, g.s.)}
&
&   \multicolumn{3}{c}{${^4_\Lambda{\rm H}}$\,($J^P=1^+$, exc)} \\
\noalign{\smallskip}
       \cline{3-5} \cline{7-9}
\noalign{\smallskip}
$\theta_{\rm lab}$  & {$q$}
&  \multicolumn{1}{c}{$\alpha\langle{d\sigma/d\Omega}\rangle^{\rm elem}_{\rm lab}$}
&  \multicolumn{1}{c}{$Z_{\rm eff}$}
&  \multicolumn{1}{r}{$d\sigma/d\Omega_{\rm lab}$} 
&
&  \multicolumn{1}{c}{$\alpha\langle{d\sigma/d\Omega}\rangle^{\rm elem}_{\rm lab}$}
&  \multicolumn{1}{c}{$Z_{\rm eff}$}
&  \multicolumn{1}{r}{$d\sigma/d\Omega_{\rm lab}$} \\
   (degree)  & {(MeV/$c$)}
&  \multicolumn{1}{c}{(mb/sr)}
&  \multicolumn{1}{c}{($\times 10^{-1}$)}
&  \multicolumn{1}{r}{($\mu$b/sr)} 
&
&  \multicolumn{1}{c}{(mb/sr)}
&  \multicolumn{1}{c}{($\times 10^{-1}$)}
&  \multicolumn{1}{r}{($\mu$b/sr)} \\
\noalign{\smallskip}\hline\noalign{\smallskip}
0   & 90  & 2.140  & 5.534 & 1184.3  &&  0.001  & 2.298 &  0.27 \\
2   & 96  & 2.003  & 5.353 & 1072.3  &&  0.004  & 2.218 &  0.98 \\
4   & 113 & 1.729  & 4.848 & 838.1   &&  0.023  & 1.996 &  4.55 \\
6   & 135 & 1.342  & 4.116 & 552.2   &&  0.062  & 1.678 & 10.5  \\
8   & 162 & 0.990  & 3.283 & 325.1   &&  0.122  & 1.321 & 16.1  \\
10  & 191 & 0.756  & 2.465 & 186.3   &&  0.181  & 0.977 & 17.7  \\
12  & 221 & 0.578  & 1.745 & 100.8   &&  0.212  & 0.680 & 14.4  \\
14  & 251 & 0.521  & 1.165 & 60.8    &&  0.249  & 0.446 & 11.1  \\
16  & 282 & 0.516  & 0.733 & 37.8    &&  0.305  & 0.275 &  8.38 \\
18  & 314 & 0.445  & 0.433 & 19.2    &&  0.358  & 0.159 &  5.68 \\
20  & 345 & 0.355  & 0.237 & 8.42    &&  0.402  & 0.085 &  3.42 \\
\end{tabular}
\end{ruledtabular}
\end{table*}

\subsection{\label{3HL}\boldmath
${^{3}_\Lambda{\rm H}}$\,($J^P=1/2^+$, g.s.) and ${^{3}_\Lambda{\rm H}}$\,($J^P=3/2^+$, exc)
}

\begin{figure}[t]
\begin{center}
  \includegraphics[width=\linewidth]{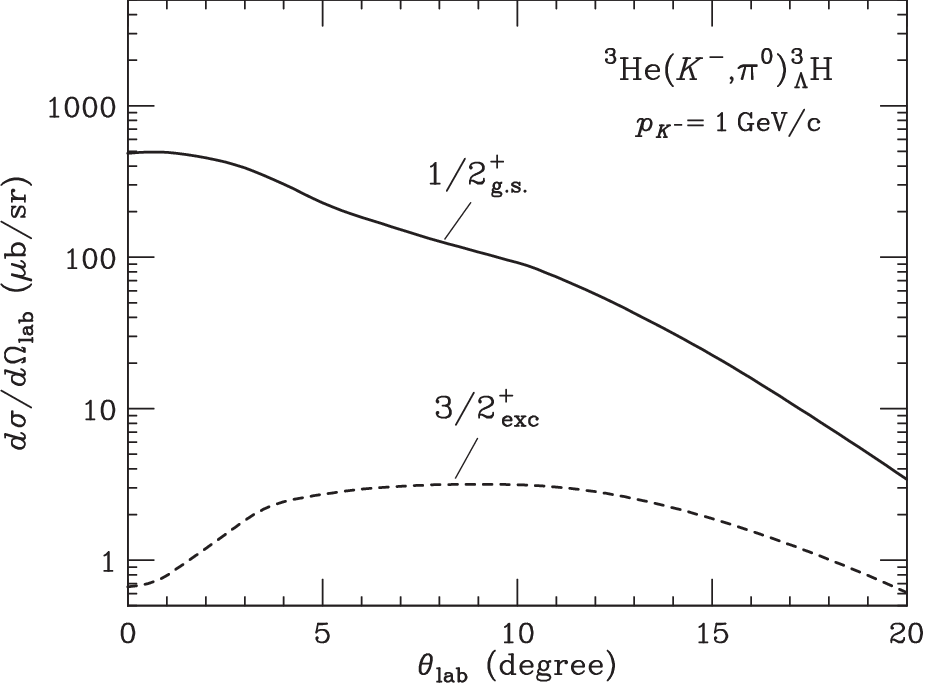}
\end{center}
\caption{\label{fig:4}
Calculated angular distributions of the laboratory differential cross sections 
$d\sigma/d\Omega_{\rm lab}$ for the 
$1/2^+_{\rm g.s.}$ and $3/2^+_{\rm exc}$ states 
in $^{3}_\Lambda$H via the $^{3}$He($K^-$,~$\pi^0$) 
reactions at $p_{K^-}=$ 1.0 GeV/$c$ in the DWIA. 
}
\end{figure}

Let us estimate numerically the production cross sections of $^{3}_\Lambda$H 
in the $^{3}$He($K^-$,~$\pi^0$) reactions at $p_{K^-}=$ 1.0 GeV/$c$ 
and $\theta_{\rm lab}=$ 0$^\circ$--20$^\circ$, where the momentum transfers 
become $q \simeq$ 80--350 MeV/$c$.
In Table~\ref{tab:3}, we list the calculated results of 
$\alpha\langle{d\sigma/d\Omega}\rangle^{\rm elem}_{\rm lab}$, $Z_{\rm eff}$,
and $d\sigma/d\Omega_{\rm lab}$ for 
${^{3}_\Lambda{\rm H}}$\,($J^P=1/2^+$, g.s.) and ${^{3}_\Lambda{\rm H}}$\,($J^P=3/2^+$, exc)
in the DWIA using the distortion parameters 
of ($\sigma_{K^-}$,~$\sigma_\pi$) = (45 mb, 32 mb).
We obtain 
$d\sigma/d\Omega_{\rm lab}(1/2^+_{\rm g.s.})=$ 484.0, 183.4, 57.1, and 7.5 $\mu$b/sr, 
and 
$d\sigma/d\Omega_{\rm lab}(3/2^+_{\rm exc})=$ 1.67, 2.94, 2.83, and 1.01 $\mu$b/sr
at $\theta_{\rm lab}=$ 0$^{\circ}$, 6$^{\circ}$, 12$^{\circ}$, and 18$^{\circ}$, respectively. 
In Fig.~\ref{fig:4}, we show the calculated angular distributions of 
$d\sigma/d\Omega_{\rm lab}$ for the $1/2^+_{\rm g.s.}$ and $3/2^+_{\rm exc}$ states 
in $^{3}_\Lambda$H via the $^{3}$He($K^-$,~$\pi^0$) reactions at $p_{K^-}=$ 1.0 GeV/$c$ 
in the DWIA.
We find that the production cross section of $d\sigma/d\Omega_{\rm lab}$ 
for the $1/2^+_{\rm g.s.}$ state is dominant at the forward angles of $\theta_{\rm lab}=$ 0$^{\circ}$--20$^{\circ}$, 
in comparison with that for $3/2^+_{\rm exc}$ state; 
we have 
$[d\sigma/d\Omega_{\rm lab}(1/2^+_{\rm g.s.})]/
[d\sigma/d\Omega_{\rm lab}(3/2^+_{\rm exc})] \simeq$ 5.6
at $\theta_{\rm lab}\simeq$ 20$^{\circ}$.

To study the feasibility of the lifetime measurements of $^{3}_\Lambda$H 
at the J-PARC experiments, 
we estimate the integrated cross sections of $\sigma_{\rm lab}$ over 
$\theta_{\rm lab}=$ 0$^{\circ}$--20$^{\circ}$, as given in Eq.~(\ref{eqn:e9}), 
in comparison with those of $^{4}_\Lambda$H. 
We find 
$\sigma_{\rm lab}(1/2^+_{\rm g.s.})=$ 25.9 $\mu$b and 
$\sigma_{\rm lab}(3/2^+_{\rm exc})=$ 0.77 $\mu$b
at $p_{K^-}=$ 1.0 GeV/$c$ in the DWIA, 
compared with $\sigma_{\rm lab}(1/2^+_{\rm g.s.})=$ 43.7 $\mu$b and 
$\sigma_{\rm lab}(3/2^+_{\rm exc})=$ 1.4 $\mu$b in the PWIA.
The distortion factor for $^3_\Lambda$H becomes $D_{\rm dis} \simeq$ 0.67--0.38 at 
$\theta_{\rm lab}=$ 0$^\circ$--20$^\circ$.
As a result, the production ratio of $\sigma_{\rm lab}(3/2^+_{\rm exc})$
to $\sigma_{\rm lab}(1/2^+_{\rm g.s.})$ amounts to 
\begin{eqnarray}
R_3=\sigma_{\rm lab}(3/2^+_{\rm exc})/\sigma_{\rm lab}(1/2^+_{\rm g.s.})
\simeq 0.030\mbox{--}0.033  \quad \mbox{for $^3_\Lambda{\rm H}$}
\label{eqn:e11}
\end{eqnarray}
at $p_{K^-}=$ 1.0 GeV/$c$. In comparison with $\sigma_{\rm lab}(1^+_{\rm exc})$ in ${^{4}_\Lambda{\rm H}}$, 
we realize that the reduction of $\sigma_{\rm lab}(3/2^+_{\rm exc})$ in ${^{3}_\Lambda{\rm H}}$ stems from 
a spread-out transition density $\rho_{\rm tr}(r)$ in Eq.~(\ref{eqn:e7}) 
due to $\langle r^2_\Lambda \rangle^{1/2}=$ 18.2 fm for the $\Lambda$ wave function 
of the $3/2^+$ state of ${^{3}_\Lambda{\rm H}}$,
whereas the absolute value of $\sigma_{\rm lab}(3/2^+_{\rm exc})$ 
depends on its pole position of the $S$ matrix for the virtual state.

The STAR Collaboration \cite{Adam20} recently reported the $\Lambda$ separation energy 
of $B_\Lambda=$ $0.41 \pm 0.12$ MeV for the $1/2^+_{\rm g.s.}$ state of ${^3_\Lambda{\rm H}}$.
To see the effects of the $\Lambda$ separation energy on the $\Lambda$ production, 
we estimate the production cross section of ${^3_\Lambda{\rm H}}$ 
in the ($K^-$,$\pi^0$) reaction at $p_{K^-}=$ 1.0 GeV/$c$, 
reproducing $B_\Lambda=$ 0.41 MeV by an additional attraction in the $\Lambda N$ interaction.
We find that the integrated cross section in the DWIA amounts to 
$\sigma_{\rm lab}(1/2^+_{\rm g.s.})=$ 41.0 $\mu$b, 
which is 60\% larger than $\sigma_{\rm lab}(1/2^+_{\rm g.s.})=$ 25.9 $\mu$b 
for $B_\Lambda=$ 0.13 MeV.
This is because the transition density $\rho_{\rm tr}(r)$ in Eq.~(\ref{eqn:e7}) becomes large, 
where the $\Lambda$ wave function of ${\varphi}^{(\Lambda)}_{0}(r)$ is shifted into the nuclear inside
due to the additional attraction, leading to $\langle r_\Lambda^2 \rangle^{1/2}=$ 6.81 fm.  
Therefore, we suggest that a precise measurement of the production cross section of 
${^{3}_\Lambda{\rm H}}$ 
gives valuable information concerning the value of $B_\Lambda$ 
with studying the nature of the $\Lambda N$ interaction and the structure of 
$^3_\Lambda$H.

\begin{table*}[bt]
\caption{\label{tab:3}
Calculated angular distributions of the laboratory differential cross sections for the 
$1/2^+_{\rm g.s.}$ and $3/2^+_{\rm exc}$ states in $^{3}_\Lambda$H via the $^{3}$He($K^-$,~$\pi^0$) 
reactions at $p_{K^-}=$ 1.0 GeV/$c$ in the DWIA. 
The distortion parameters of ($\sigma_{K^-}$,~$\sigma_\pi$) = (45 mb, 32 mb) are used.
}
\begin{ruledtabular}
\begin{tabular}{ccccrcccr}
&
&   \multicolumn{3}{c}{${^3_\Lambda{\rm H}}$\,($J^P=1/2^+$, {\rm g.s.})}
&
&   \multicolumn{3}{c}{${^3_\Lambda{\rm H}}$\,($J^P=3/2^+$, {\rm exc})} \\
\noalign{\smallskip}
       \cline{3-5} \cline{7-9}
\noalign{\smallskip}
$\theta_{\rm lab}$  &  {$q$}
&  \multicolumn{1}{c}{$\alpha\langle{d\sigma/d\Omega}\rangle^{\rm elem}_{\rm lab}$}
&  \multicolumn{1}{c}{$Z_{\rm eff}$}
&  \multicolumn{1}{r}{$d\sigma/d\Omega_{\rm lab}$} 
&
&  \multicolumn{1}{c}{$\alpha\langle{d\sigma/d\Omega}\rangle^{\rm elem}_{\rm lab}$}
&  \multicolumn{1}{c}{$Z_{\rm eff}$}
&  \multicolumn{1}{r}{$d\sigma/d\Omega_{\rm lab}$} \\
   (degree) &  {(MeV/$c$)}
&  \multicolumn{1}{c}{(mb/sr)}
&  \multicolumn{1}{c}{($\times 10^{-1}$)}
&  \multicolumn{1}{r}{($\mu$b/sr)} 
&
&  \multicolumn{1}{c}{(mb/sr)}
&  \multicolumn{1}{c}{($\times 10^{-1}$)}
&  \multicolumn{1}{r}{($\mu$b/sr)} \\
\noalign{\smallskip}\hline\noalign{\smallskip}
0   & 79  & 1.488  & 3.253  & 484.0  && 0.010  & 0.668 & 0.67 \\
2   & 86  & 1.465  & 3.090  & 452.8  && 0.019  & 0.626 & 1.19 \\
4   & 104 & 1.141  & 2.664  & 304.1  && 0.047  & 0.520 & 2.43 \\
6   & 129 & 0.869  & 2.110  & 183.4  && 0.075  & 0.390 & 2.94 \\
8   & 157 & 0.815  & 1.560  & 127.2  && 0.116  & 0.271 & 3.15 \\
10  & 187 & 0.842  & 1.092  & 92.0   && 0.177  & 0.178 & 3.14 \\
12  & 218 & 0.780  & 0.733  & 57.1   && 0.253  & 0.112 & 2.83 \\
14  & 249 & 0.662  & 0.474  & 31.4   && 0.326  & 0.068 & 2.21 \\
16  & 281 & 0.533  & 0.298  & 15.9   && 0.390  & 0.040 & 1.56 \\
18  & 312 & 0.411  & 0.183  & 7.50   && 0.441  & 0.023 & 1.01 \\
20  & 344 & 0.314  & 0.109  & 3.41   && 0.480  & 0.013 & 0.61 \\
\end{tabular}
\end{ruledtabular}
\end{table*}

\subsection{Recoil effects}

We have recognized that the recoil effects are essential in production reactions 
on the very light nuclear target such as $^{4}$He \cite{Harada19}. 
Thus, the recoil correction may significantly enlarge the production cross sections
because the {\it effective} momentum transfers denote 
$q_{\rm eff}\simeq$ 51--223 MeV/$c$ at $\theta_{\rm lab}=$ 0$^\circ$--20$^\circ$
with $M_C/M_D=0.647$ for ${^3_\Lambda{\rm H}}$, 
which achieve the $\Lambda$ production in the near-recoilless reaction, rather than 
$q_{\rm eff}\simeq$ 66--253 MeV/$c$ with $M_C/M_D=0.734$ for ${^4_\Lambda{\rm H}}$. 
When the recoil correction is omitted ($M_C/M_D \to 1$), 
the integrated cross sections amount to 
$\sigma_{\rm lab}({^3_\Lambda{\rm H}}; 1/2^+_{\rm g.s.})=$ 9.74 $\mu$b
and $\sigma_{\rm lab}({^4_\Lambda{\rm H}}; 0^+_{\rm g.s.})=$ 35.2 $\mu$b in the DWIA, 
and 
$\sigma_{\rm lab}({^3_\Lambda{\rm H}}; 1/2^+_{\rm g.s.})=$ 17.9 $\mu$b
and $\sigma_{\rm lab}({^4_\Lambda{\rm H}}; 0^+_{\rm g.s.})=$ 89.9 $\mu$b in the PWIA. 
The values of $\sigma_{\rm lab}$ with the recoil effects are larger than 
those of $\sigma_{\rm lab}$ without the recoil effects 
by a factor of about 2.5 (1.7) for ${^3_\Lambda{\rm H}}$ (${^4_\Lambda{\rm H}}$).
Therefore, we confirm the benefit of the use of the $^{3,4}$He targets 
in the $\Lambda$ production via the ($K^-$,~$\pi^0$) reaction. 

\begin{table*}[bt]
\caption{\label{tab:4}
Calculated angular distributions of the laboratory differential cross sections for  
${^{3,4}_\Lambda{\rm H}}$ in the $^{3,4}$He($\pi^-$,~$K^0$) reactions
at $p_{\pi^-}=$ 1.05 GeV/$c$ in the DWIA. 
The distortion parameters of ($\sigma_\pi$,~$\sigma_{K^+}$) = (30 mb, 15 mb) are used.
}
\begin{ruledtabular}
\begin{tabular}{ccccrccccr}
&   \multicolumn{4}{c}{${^3_\Lambda{\rm H}}$\,($J^P=1/2^+$, {\rm g.s.})} 
&
&   \multicolumn{4}{c}{${^4_\Lambda{\rm H}}$\,($J^P=0^+$, {\rm g.s.})} \\
\noalign{\smallskip}
       \cline{2-5} \cline{7-10}
\noalign{\smallskip}
$\theta_{\rm lab}$
&  \multicolumn{1}{c}{$q$}
&  \multicolumn{1}{c}{$\alpha\langle{d\sigma/d\Omega}\rangle^{\rm elem}_{\rm lab}$}
&  \multicolumn{1}{c}{$Z_{\rm eff}$}
&  \multicolumn{1}{c}{$d\sigma/d\Omega_{\rm lab}$} 
&
&  \multicolumn{1}{c}{$q$}
&  \multicolumn{1}{c}{$\alpha\langle{d\sigma/d\Omega}\rangle^{\rm elem}_{\rm lab}$}
&  \multicolumn{1}{c}{$Z_{\rm eff}$}
&  \multicolumn{1}{c}{$d\sigma/d\Omega_{\rm lab}$} \\
   (degree)
&  \multicolumn{1}{c}{(MeV/$c$)}
&  \multicolumn{1}{c}{($\mu$b/sr)}
&  \multicolumn{1}{r}{($\times 10^{-2}$)}
&  \multicolumn{1}{r}{($\mu$b/sr)} 
&
&  \multicolumn{1}{c}{(MeV/$c$)}
&  \multicolumn{1}{c}{($\mu$b/sr)}
&  \multicolumn{1}{r}{($\times 10^{-2}$)}
&  \multicolumn{1}{r}{($\mu$b/sr)} \\
\noalign{\smallskip}\hline\noalign{\smallskip}
 0  & 351 & 570.7  & 1.276  & 7.28  && 362 & 624.0 & 3.015 & 18.82  \\
 2  & 352 & 567.2  & 1.247  & 7.07  && 364 & 618.7 & 2.938 & 18.18  \\
 4  & 357 & 556.7  & 1.164  & 6.48  && 368 & 603.3 & 2.719 & 16.40  \\
 6  & 364 & 539.4  & 1.040  & 5.61  && 375 & 579.0 & 2.387 & 13.82  \\ 
 8  & 374 & 515.7  & 0.890  & 4.59  && 384 & 547.7 & 1.988 & 10.89  \\
10  & 386 & 486.3  & 0.732  & 3.56  && 395 & 511.0 & 1.568 &  8.01  \\
12  & 400 & 452.0  & 0.579  & 2.62  && 408 & 470.7 & 1.168 &  5.50  \\
14  & 416 & 413.9  & 0.443  & 1.83  && 423 & 428.2 & 0.819 &  3.51  \\
16  & 434 & 373.5  & 0.327  & 1.22  && 440 & 384.6 & 0.536 &  2.06  \\
18  & 453 & 332.3  & 0.234  & 0.78  && 458 & 341.1 & 0.324 &  1.10  \\
20  & 473 & 291.7  & 0.163  & 0.47  && 477 & 298.7 & 0.176 &  0.53  \\
\end{tabular}
\end{ruledtabular}
\end{table*}

\subsection{\boldmath
Comparison with $^{3,4}$He($\pi^-$, $K^0$) reactions}

It is also interesting to discuss the production cross sections of $^{3,4}_\Lambda{\rm H}$ 
in the endothermic ($\pi^-$,~$K^0$) reactions on $^{3,4}{\rm He}$ at 
$p_{\pi^-}=$ 1.05 GeV/$c$ and $\theta_{\rm lab}=$ 0$^\circ$--20$^\circ$, where the
high momentum transfers of $q \simeq$ 350--500 MeV/$c$ are expected to bring  
benefits to the use of the $^{3,4}$He targets \cite{Harada19,Harada19a}. 
In Table~\ref{tab:4},
we list the calculated results of 
$\alpha\langle{d\sigma/d\Omega}\rangle^{\rm elem}_{\rm lab}=\alpha|\overline{f}_{\pi^-p \to K^0\Lambda}|^2$, 
$Z_{\rm eff}$, and $d\sigma/d\Omega_{\rm lab}$ for 
${^3_\Lambda{\rm H}}$\,($J^P=1/2^+$, {\rm g.s.}) and 
${^4_\Lambda{\rm H}}$\,($J^P=0^+$, {\rm g.s.}) 
at 1.05 GeV/$c$ in the DWIA.
We find the updated values of 
$d\sigma/d\Omega_{\rm lab}({^3_\Lambda{\rm H}}; 1/2^+_{\rm g.s.})=$ 7.28, 5.61, 2.62, and 0.78 $\mu$b/sr 
and 
$d\sigma/d\Omega_{\rm lab}({^4_\Lambda{\rm H}}; 0^+_{\rm g.s.})=$ 18.82, 13.82, 5.50, and 1.10 $\mu$b/sr 
at $\theta_{\rm lab}=$ 0$^{\circ}$, 6$^{\circ}$, 12$^{\circ}$, and 18$^{\circ}$, respectively, 
compared with the previous works \cite{Harada19,Harada19a}.
These results lead to $\sigma_{\rm lab}({^{3}_\Lambda{\rm H}}; 1/2^+_{\rm g.s.})=$ 0.93 $\mu$b and 
$\sigma_{\rm lab}({^{4}_\Lambda{\rm H}}; 0^+_{\rm g.s.})=$ 2.02 $\mu$b. 
Note that the angular dependences of 
$\alpha|\overline{f}_{\pi^-p \to K^0\Lambda}|^2$ 
for ${^4_\Lambda{\rm H}}$ and ${^3_\Lambda{\rm H}}$ are 
very similar in the ($K^-$, $\pi^0$) reactions at $\theta_{\rm lab}=$ 0$^\circ$--20$^\circ$, 
whereas the absolute values of the former are slightly larger than those of the latter.

Moreover, we find that 
the values of $d\sigma/d\Omega_{\rm lab}({^{3}_\Lambda{\rm H}}; 1/2^+_{\rm g.s.})$ in the region of 
$q >$ 350 MeV/$c$ are enhanced by more than 14\% owing to the use of the CDCC wave functions 
for $^3_\Lambda$H in our calculations, in comparison with those obtained by omitting the couplings 
between $[d\otimes\Lambda]$ and $[(d^{*})_n \otimes\Lambda]$ channels in the CDCC, 
where $(d^{*})_n$ denote the $n$-th continuum-discretized excited states of the deuteron core nucleus. 
This implies that the excited-state components of $(d^{*})_n$ contribute 
to the $^3_\Lambda$H production \cite{Harada15a}, so its production yield grows with increasing $q$.

\begin{figure}[t]
\begin{center}
  \includegraphics[width=\linewidth]{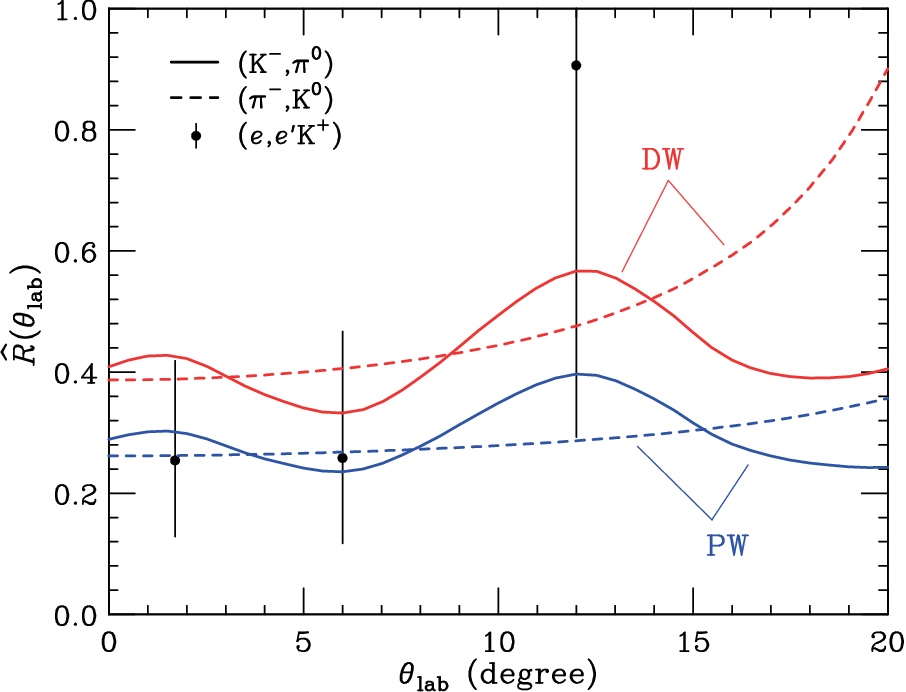}
\end{center}
\caption{\label{fig:5}
Comparison among the ratios of 
$\hat{R}(\theta_{\rm lab})=[d\sigma/d\Omega_{\rm lab}({^{3}_\Lambda{\rm H}})]/
  [d\sigma/d\Omega_{\rm lab}({^{4}_\Lambda{\rm H}})]$ in the DW and the PW, 
as a function of $\theta_{\rm lab}$.
Solid and dashed curves denote the calculated values in the ($K^-$,~$\pi^0$) reaction 
at 1.0 GeV/$c$ and the ($\pi^-$,~$K^0$) reaction at 1.05 GeV/$c$, respectively.
The experimental data in the ($e$,~$e'$$K^+$) reaction at the virtual 
photon $\gamma^*$ mass $Q^2=$ 3.5 GeV$^2$ are taken from Ref.~\cite{Dohrmann04}.  
}
\end{figure}

\subsection{\boldmath
$^{3}_\Lambda{\rm H}$ v.s. $^{4}_\Lambda{\rm H}$}

To compare the production cross sections between $^{3}_\Lambda{\rm H}$ 
and $^{4}_\Lambda{\rm H}$, 
we consider the ratio of $^3_\Lambda{\rm H}$ to $^4_\Lambda{\rm H}$ on 
the angular distributions of $d\sigma/d\Omega_{\rm lab}$,
\begin{eqnarray}
\hat{R}(\theta_{\rm lab})=[d\sigma/d\Omega_{\rm lab}({^{3}_\Lambda{\rm H}})]/
  [d\sigma/d\Omega_{\rm lab}({^{4}_\Lambda{\rm H}})]. 
\label{eqn:e12}
\end{eqnarray}
Here we use only the production cross sections of the ${^{3,4}_\Lambda{\rm H}}$ ground states 
because the contributions of the ${^{3,4}_\Lambda{\rm H}}$ excited states to the $\Lambda$ productions are very small, 
as discussed above. 
In Fig.~\ref{fig:5}, we show the calculated values of $\hat{R}(\theta_{\rm lab})$ 
in the ($K^-$,~$\pi^0$) reaction at 1.0 GeV/$c$, together with those of 
$\hat{R}(\theta_{\rm lab})$ in the ($\pi^-$,~$K^0$) reaction at 1.05 GeV/$c$.
In the ($K^-$,~$\pi^0$) reaction at 1.0 GeV/$c$, 
we find that
the values of $\hat{R}(\theta_{\rm lab})$ fluctuate in the range of 
0.3--0.6 at $\theta_{\rm lab}=$ 0$^\circ$--20$^\circ$.
This behavior mainly indicates the difference
in $\alpha\langle{d\sigma/d\Omega}\rangle^{\rm elem}_{\rm lab}$ between $^3_\Lambda$H and $^4_\Lambda$H, 
rather than the angular dependence of $Z_{\rm eff}$ at $\theta_{\rm lab}=$ 0$^\circ$--20$^\circ$
that correspond to $q \simeq$ 80--350 MeV/$c$. 
On the other hand, in the ($\pi^-$,~$K^0$) reaction at 1.05 GeV/$c$, 
we find $\hat{R}(\theta_{\rm lab}) \simeq$ 0.4--0.8 at $\theta_{\rm lab}=$ 0$^\circ$--20$^\circ$
that correspond to $q \simeq$ 350--470 MeV/$c$. 
This behavior indicates the angular dependence of 
$Z_{\rm eff}({^3_\Lambda{\rm H}_{\rm g.s.}})/Z_{\rm eff}({^4_\Lambda{\rm H}_{\rm g.s.}})$, 
which is related to the $A=$ 3, 4 form factors $F(q)$ over the $\Lambda$ production processes
because the angular dependences of $\alpha\langle{d\sigma/d\Omega}\rangle^{\rm elem}_{\rm lab}$ for $^3_\Lambda$H 
and $^4_\Lambda$H are very similar.
In Fig.~{\ref{fig:5}}, we also draw the experimental data taken from 
the $^{3,4}$He($e$,~$e'$$K^+$) reaction at the virtual photon 
$\gamma^*$ mass $Q^2=$ 3.5 GeV$^2$~\cite{Dohrmann04,Mart08}.
It seems that the calculated results of $\hat{R}(\theta_{\rm lab})$ in the ($\pi^-$,~$K^0$) reaction 
can simulate the data of the ($e$,~$e'$$K^+$) reaction 
because the values of $q$ for the former and the latter are roughly the same. 
Moreover, we estimate the ratio of $\sigma_{\rm lab}({^{3}_\Lambda{\rm H}})$ 
to $\sigma_{\rm lab}({^{4}_\Lambda{\rm H}})$ 
on the integrated cross sections over $\theta_{\rm lab}=$ 0$^\circ$--20$^\circ$, 
which is given by 
\begin{eqnarray}
R_{34}=\sigma_{\rm lab}({^{3}_\Lambda{\rm H}})/\sigma_{\rm lab}({^{4}_\Lambda{\rm H}}).
\label{eqn:e13}
\end{eqnarray}
In the ($K^-$,~$\pi^0$) reaction at 1.0 GeV/$c$, 
we find $R_{34}=$ 0.41 in the DWIA.
Considering some ambiguities in our eikonal-DWIA calculations, 
we also find $R_{34}=$ 0.29 in the PWIA, omitting the distortions. 
Consequently, we have $R_{34} \simeq$ 0.3--0.4.
Note that the value of $R_{34}$ depends on $B_\Lambda$ for ${^3_\Lambda{\rm H}}$; 
we find $R_{34} \simeq$ 0.65 when we use $B_\Lambda=$ 0.41 MeV, 
as discussed in Sect.~\ref{3HL}. 
It strongly suggests that the production of ${^{3}_\Lambda{\rm H}}$ 
is a promising subject to be observed experimentally 
based on a successive measurement of ${^{4}_\Lambda{\rm H}}$ at the J-PARC experiment. 
In the ($\pi^-$,~$K^0$) reaction at 1.05 GeV/$c$, 
we find $R_{34}=$ 0.46 in the DWIA and $R_{34}=$ 0.28 in the PWIA, 
leading to $R_{34} \simeq$ 0.3--0.4.
The comparison between the nuclear ($K^-$,~$\pi^0$) and ($\pi^-$,~$K^0$) reactions 
on $\hat{R}(\theta_{\rm lab})$ provides a better understanding of not only the structure of 
the $^{3,4}_\Lambda$H bound states but also the production mechanism of these states.

\section{Summary and conclusion}
\label{sect:6}
We have investigated theoretically productions of $^{3,4}_\Lambda$H bound states 
via $^{3,4}$He($K^-$,~$\pi^0$) reactions
in the DWIA with the optimal Fermi-averaging $K^-p\to\pi^0\Lambda$ $t$ matrix. 
We have calculated the laboratory differential cross sections of 
$d\sigma/d\Omega_{\rm lab}$ and the integrated cross sections 
of $\sigma_{\rm lab}$ by the non-spin-flip $\Delta S= 0$ production 
in the ${^{3,4}{\rm He}}$($K^-$,~$\pi^0$) reactions at 1.0 GeV/$c$ and $\theta_{\rm lab}=$ 0$^\circ$--20$^\circ$, 
together with those by the spin-flip $\Delta S= 1$ production.
We have also compared these cross sections with those in the ($\pi^-$,~$K^0$) reactions at 1.05 GeV/$c$. 
The results are summarized as follows:
\begin{itemize}
\item[(i)] 
The calculated integrated cross sections of the $0^+_{\rm g.s.}$ and $1^+_{\rm exc}$ states 
of ${^{4}_\Lambda{\rm H}}$ amount to 
$\sigma_{\rm lab}(0^+_{\rm g.s.})=$ 63.1 $\mu$b and 
$\sigma_{\rm lab}(1^+_{\rm exc})=$ 3.8 $\mu$b, respectively, leading to 
$R_4=\sigma_{\rm lab}(1^+_{\rm exc})/\sigma_{\rm lab}(0^+_{\rm g.s.}) \simeq 0.06\mbox{--}0.07$. 
The production of the $0^+_{\rm g.s.}$ state dominates in the forward angles of 
$\theta_{\rm lab}=$ 0$^\circ$--20$^\circ$, in comparison with that of the $1^+_{\rm g.s.}$ state.

\item[(ii)] 
The calculated integrated cross sections of the $1/2^+_{\rm g.s.}$ and $3/2^+_{\rm exc}$ states 
of ${^{3}_\Lambda{\rm H}}$ amount to 
$\sigma_{\rm lab}(1/2^+_{\rm g.s.})=$ 25.9 $\mu$b and 
$\sigma_{\rm lab}(3/2^+_{\rm exc})=$ 0.77 $\mu$b, respectively.
This leads to $R_3=\sigma_{\rm lab}(3/2^+_{\rm exc})/\sigma_{\rm lab}(1/2^+_{\rm g.s.})
\simeq 0.03$, of which value is a half as large as that of $R_4$.

\item[(iii)]
The calculated angular distributions of the in-medium $K^-p\to \pi^0\Lambda$ 
differential cross sections $\alpha|\overline{f}_{\pi^0\Lambda}|^2$ 
for ${^{3}_\Lambda{\rm H}}$ are remarkably different from those for ${^{4}_\Lambda{\rm H}}$, 
caused by the optimal Fermi-averaging in the nuclear ($K^-$, $\pi^0$) reactions, 
whereas $\alpha |\overline{g}_{\pi^0\Lambda}|^2$ for ${^{3,4}_\Lambda{\rm H}}$ 
are very similar to each other.

\item[(iv)] 
The recoil effects are important in productions of $^{3,4}_\Lambda$H 
owing to the benefit of the use of the $^{3,4}$He targets via 
the nuclear ($K^-$, $\pi^0$) reactions, as well as the 
nuclear ($\pi^-$,~$K^0$) reactions.

\end{itemize}
\noindent
In conclusion, we show that 
the comparison in $d\sigma/d\Omega_{\rm lab}$ and $\sigma_{\rm lab}$ 
between $^{4}_\Lambda{\rm H}$ and $^{3}_\Lambda{\rm H}$
provides examining the mechanism of the production and structure of $^{3,4}_\Lambda$H 
in the ($K^-$,~$\pi^0$) reactions on $^{3,4}$He at $p_{K^-}=$ 1.0 GeV/$c$; 
the calculated results indicate $R_{34}=\sigma_{\rm lab}(^{3}_\Lambda{\rm H})
/\sigma_{\rm lab}(^{4}_\Lambda{\rm H})\simeq$ 0.3--0.4.
This investigation confirms the feasibility of the lifetime 
measurements of $^3_\Lambda$H at the J-PARC experiments.

\begin{acknowledgments}
The authors thank Dr.~Y.~Ma and Dr.~F.~Sakuma for many valuable discussions. 
This work was supported by Grants-in-Aid for
Scientific Research (KAKENHI) from the Japan Society for
the Promotion of Science: 
Scientific Research (C) (Grant No.~JP20K03954).
\end{acknowledgments}

\appendix

\section{\boldmath
Explicit forms of the differential cross sections
}
\label{app:1}

To consider the laboratory differential cross sections of the $A$($K^-$, $\pi$)$B$ reaction in the DWIA, 
we will define wave functions of the initial and final states, ${\Psi}_A$ and ${\Psi}_B$, 
in the $jj$ coupling scheme:
\begin{eqnarray}
&& \vert {\Psi}_A \rangle 
   =   \hat{\cal A} \left[ 
  \Phi_{J_C}\otimes\phi^{(N)}_{(\ell_1{1 \over 2})j_1}
  \right]_{J_A}^{M_A}, \\
&& \vert {\Psi}_B \rangle =
  \sum_{J_Cj_2} \left[ 
  \Phi_{J_C}\otimes\phi^{(\Lambda)}_{(\ell_2{1 \over 2})j_2}
  \right]_{J_B}^{M_B}, 
\label{eqn:a1}
\end{eqnarray}
where $\Phi_{J_C}$, $\phi^{(N)}_{(\ell_1{1 \over 2})j_1}$, and 
$\phi^{(\Lambda)}_{(\ell_2{1 \over 2})j_2}$ are 
wave functions of a core nucleus, a nucleon in the target nucleus $A$, 
and $\Lambda$ in the hypernucleus $B$, respectively. 
$\hat{\cal A}$ is the antisymmetrized operator for nucleons. 
The meson distorted waves for outgoing $\pi$ and incoming $K^-$ are 
written by the partial wave expansion
\begin{eqnarray}
\chi^{(-)*}_{\pi}({\bm p}_{\pi},{\bm r})\chi^{(+)}_{K^-}({\bm p}_{K^-},{\bm r})
&=&\sum_{\ell m} \sqrt{4 \pi[\ell]}\, i^\ell \,
\widetilde{j}_{\ell m}(\theta_{\rm lab},r)Y_{\ell m}(\hat{\bm r}), 
\label{eqn:a2}
\end{eqnarray}
where $\widetilde{j}_{\ell m}(\theta_{\rm lab},r)$ 
is the radial distorted wave with the angular momentum 
with ($\ell$,~$m$),  
and $\theta_{\rm lab}$ is the scattering angle to the forward 
direction in the nuclear ($K^-$, $\pi$) reaction. 

The explicit form of the differential cross section with the non-spin-flip $\Delta S=0$ 
processes in Eq.~(\ref{eqn:e1}) is written as
\begin{eqnarray}
   \left(\frac{d\sigma}{d\Omega}\right)^{J^P_B (\Delta S=0)}_{\rm lab} 
& = & \alpha
    \sum_{\ell m cc'}
    (S_{c}^{1/2})^*(S_{c'}^{1/2})
    \overline{f}_{\pi\Lambda}^{*} \overline{f}_{\pi\Lambda}
    (-)^{2J_B+J_{C}+J_{C}'-1}
\nonumber \\
&&  \times
     [J_B][\ell] 
      \sqrt{[J_C][j_2][j_1][J_{C}'][j_2'][j_1']} \, 
     [I^{m}_{j_2j_1\ell}(\theta_{\rm lab})]^*
     [I^{m}_{j_2'j_1'\ell}(\theta_{\rm lab})] 
\nonumber \\
&& \times \left( \begin{array}{ccc} 
                 j_2 &    \ell  &  j_1  \\
         -\frac{1}{2}&       0  &  \frac{1}{2}  \\
                 \end{array}  \right)  
    \left\{ \begin{array}{ccc} 
                  J_B   &   J_A  &  \ell  \\
                  j_1   &   j_2  &  J_C \\
    \end{array}  \right\} \frac{[1+(-)^{\ell_2+\ell_1+\ell}]}{2}
\nonumber \\
&& \times \left( \begin{array}{ccc} 
                 j_2 &    \ell  &  j_1  \\
         -\frac{1}{2}&       0  &  \frac{1}{2}  \\
                 \end{array}  \right)  
    \left\{ \begin{array}{ccc} 
                  J_B   &   J_A  &  \ell  \\
                  j_1'   &   j_2'  &  J_C' \\
    \end{array}  \right\} \frac{[1+(-)^{\ell_2'+\ell_1'+\ell}]}{2}
\label{eqn:a3}
\end{eqnarray}
with the integral 
\begin{eqnarray}
&& 
I^{m}_{j_2j_1\ell}(\theta_{\rm lab})
= \int_0^{\infty} dr r^2
     \varphi^{(\Lambda)*}_{j_2}(r)
     \widetilde{j}_{\ell m}(\theta_{\rm lab},{M_C \over M_D}r) 
     \varphi^{(N)}_{j_1}(r), 
\label{eqn:a2}
\end{eqnarray}
where $S_{c}^{1/2}$ is the spectroscopic amplitude for channel $c=\{J_C j_1j_2\ell_1\ell_2\}$
that may involve generally the isospin states, 
and the kinematical factor $\alpha$ arising from a translation from a two-body meson-nucleon 
laboratory system to a meson-nucleus laboratory system \cite{Dover83} is given by 
\begin{equation}
 \alpha=
 {p_{\pi} E_{\pi} \over p^{(0)}_{\pi} E^{(0)}_{\pi}}
 \biggl(1+ {E^{(0)}_{\pi} \over E^{(0)}_{B}}
        {{p^{(0)}_{\pi} - p^{(0)}_{K} \cos\theta_{\rm lab}}
        \over p^{(0)}_{\pi}} \biggr)
        \left(1+{E_{\pi} \over E_B}{p_{\pi}-p_{K} \cos{\theta_{\rm lab}} \over p_{\pi}}\right)^{-1}, 
\label{eqn:a4}
\end{equation}
where 
$p^{(0)}_{K}$ and $p^{(0)}_{\pi}$ ($E^{(0)}_{\pi}$ and 
$E^{(0)}_{B}$) are 
laboratory momenta of $K^-$ and $\pi$ (laboratory energies of $\pi$ and $\Lambda$)
in the two-body $K^- N\to \pi\Lambda$ reaction, respectively.

For the spin-flip $\Delta S=1$ processes, substituting the relation 
\begin{eqnarray}
Y_{\ell m}\,{\bm \sigma} \cdot {\hat{\bm n}}
&=& Y_{\ell m}\sigma_y = \frac{i}{\sqrt{2}}
\sum_{\mu=-1}^{1}\sum_J \langle \ell m 1 \mu | J M \rangle[Y_\ell \otimes \sigma_1 ]^M_J
\label{eqn:a5}
\end{eqnarray} 
into Eq.~(\ref{eqn:e1}) and using the Racah algebra, 
we have the differential cross section with $\Delta S=1$, which is written by 
\begin{eqnarray}
   \left(\frac{d\sigma}{d\Omega}\right)^{J^P_B (\Delta S=1)}_{\rm lab} 
& = & \alpha
    \sum_{J M cc'}\sum_{\ell m \ell' m'}\sum_{\mu \mu'}
    (S_{c}^{1/2})^*(S_{c'}^{1/2})
    \overline{g}_{\pi\Lambda}^{*} \overline{g}_{\pi\Lambda} 
\nonumber \\
&&  \times
    \frac{1}{2}\langle \ell m 1 \mu | J M \rangle\langle J M | \ell' m' 1 \mu' \rangle 
    (-)^{2J_B+J_{C}+J_{C}'+2J+j_1+j_1'+\ell_2+\ell_2'}
\nonumber \\
&&  \times
     6\,[J_B][\ell][\ell'] 
     \sqrt{[J_C][j_2][j_1][\ell_2][\ell_1][J_{C}'][j_2'][j_1'][\ell_2'][\ell_1']}
\nonumber \\
&&  \times
     [I^{m}_{j_2j_1\ell}(\theta_{\rm lab})]^*
     [I^{m}_{j_2'j_1'\ell'}(\theta_{\rm lab})] 
\nonumber \\
&& \times 
    \left\{ \begin{array}{ccc} 
                  J_B   &   J_A  &  J  \\
                  j_1   &   j_2  &  J_C \\
    \end{array}  \right\} 
    \left( \begin{array}{ccc} 
                 \ell_2 &    \ell  &  \ell_1  \\
                      0  &      0  &       0  \\
                 \end{array}  \right)  
    \left\{ \begin{array}{ccc} 
                  \ell_2   &  \frac{1}{2} &  j_2 \\
                  \ell_1   &  \frac{1}{2} &  j_1 \\
                  \ell     &        1     &  J   \\
    \end{array}  \right\} 
\nonumber \\
&& \times 
    \left\{ \begin{array}{ccc} 
                  J_B   &   J_A  &  J  \\
                  j_1'   &   j_2'  &  J_C' \\
    \end{array}  \right\} 
    \left( \begin{array}{ccc} 
                 \ell_2' &    \ell'  &  \ell_1'  \\
                      0  &      0  &       0  \\
                 \end{array}  \right)  
    \left\{ \begin{array}{ccc} 
                  \ell_2'   &  \frac{1}{2} &  j_2' \\
                  \ell_1'   &  \frac{1}{2} &  j_1' \\
                  \ell'     &        1     &  J   \\
    \end{array}  \right\}.
\label{eqn:a6}
\end{eqnarray}


\clearpage

\end{document}